\documentclass[reprint, coublecolumn, superscriptaddress, prb]{revtex4-1}
\usepackage{lineno}
\usepackage{graphicx}
\usepackage{float}
\usepackage{amssymb}
\usepackage{mathrsfs}
\usepackage{hyperref}
\usepackage[british]{babel}

\begin{document}

\title{Violating Bell's inequality with remotely-connected superconducting qubits}

\author{Y. P. Zhong}
\affiliation{Institute for Molecular Engineering, University of Chicago, Chicago IL 60637, USA}
\author{H.-S. Chang}
\affiliation{Institute for Molecular Engineering, University of Chicago, Chicago IL 60637, USA}
\author{K. J. Satzinger}
\affiliation{Institute for Molecular Engineering, University of Chicago, Chicago IL 60637, USA}
\affiliation{Department of Physics, University of California, Santa Barbara, CA 93106, USA}
\author{M.-H. Chou}
\affiliation{Institute for Molecular Engineering, University of Chicago, Chicago IL 60637, USA}
\affiliation{Department of Physics, University of Chicago, Chicago IL 60637, USA}
\author{A. Bienfait}
\affiliation{Institute for Molecular Engineering, University of Chicago, Chicago IL 60637, USA}
\author{C. R. Conner}
\affiliation{Institute for Molecular Engineering, University of Chicago, Chicago IL 60637, USA}
\author{\'E. Dumur}
\affiliation{Institute for Molecular Engineering, University of Chicago, Chicago IL 60637, USA}
\affiliation{Institute for Molecular Engineering and Materials Science Division, Argonne National Laboratory, Argonne IL 60439, USA}
\author{J. Grebel}
\affiliation{Institute for Molecular Engineering, University of Chicago, Chicago IL 60637, USA}
\author{G. A. Peairs}
\affiliation{Institute for Molecular Engineering, University of Chicago, Chicago IL 60637, USA}
\affiliation{Department of Physics, University of California, Santa Barbara, CA 93106, USA}
\author{R. G. Povey}
\affiliation{Institute for Molecular Engineering, University of Chicago, Chicago IL 60637, USA}
\affiliation{Department of Physics, University of Chicago, Chicago IL 60637, USA}
\author{D. I. Schuster}
\affiliation{Department of Physics, University of Chicago, Chicago IL 60637, USA}
\author{A. N. Cleland}
\affiliation{Institute for Molecular Engineering, University of Chicago, Chicago IL 60637, USA}
\affiliation{Institute for Molecular Engineering and Materials Science Division, Argonne National Laboratory, Argonne IL 60439, USA}

\begin{abstract} 
 Quantum communication relies on the efficient generation of entanglement between remote quantum nodes, due to entanglement's key role in achieving and verifying secure communications \cite{Cirac1997}. Remote entanglement has been realized using a number of different probabilistic schemes \cite{Narla2016,Dickel2018}, but deterministic remote entanglement has only recently been demonstrated, using a variety of superconducting circuit approaches \cite{Kurpiers2018,Axline2018,Campagne2018}. However, the deterministic violation of a Bell inequality \cite{Bell1964}, a strong measure of quantum correlation, has not to date been demonstrated in a superconducting quantum communication architecture, in part because achieving sufficiently strong correlation requires fast and accurate control of the emission and capture of the entangling photons. Here we present a simple and robust architecture for achieving this benchmark result in a superconducting system.
 \end{abstract}

\maketitle

Superconducting quantum circuits have made significant progress over the past few years, demonstrating improved qubit lifetimes, higher gate fidelities, and increasing circuit complexity \cite{Devoret2013,Kelly2015}. Superconducting qubits also offer highly flexible quantum control over other systems, including electromagnetic \cite{Hofheinz2009,Vlastakis2013} and mechanical resonators \cite{OConnell2010,Chu2017}. These devices are thus appealing for testing quantum communication protocols, with recent demonstrations of deterministic remote state transfer and entanglement generation \cite{Kurpiers2018,Axline2018,Campagne2018}. The Bell inequality \cite{Bell1964} is an important benchmark for entanglement, providing a straightforward test of whether a local and deterministic theory can explain measured correlations.  To date, however, only local violations of the Bell or Leggett-Garg \cite{leggett1985} inequalities have been demonstrated using superconducting qubits \cite{Ansmann2009, Palacios2010}, as remote state transfer and entanglement generation with sufficiently high fidelity is still an experimental challenge.

Here we present two distinct methods that violate the Clauser-Horne-Shimony-Holt (CHSH) \cite{Clauser1969} form of the Bell inequality, using a pair of superconducting qubits coupled through a 78 cm-long transmission line, with the photon emission and capture rates controlled by a pair of electrically-tunable couplers \cite{Chen2014}. In one experiment, we use a single standing mode of the transmission line to relay quantum states between the qubits, achieving a transfer fidelity of $0.952 \pm 0.009$. This enables the deterministic generation of a Bell state with a fidelity of $0.957 \pm 0.005$. Measurements on this remotely-entangled Bell state achieve a CHSH correlation $S = 2.237 \pm 0.036$, exceeding the classical correlation limit of $|S|\le 2$ by 6.6 standard deviations. In the second experiment, we control the time-dependent emission and capture rates of itinerant photons through the transmission line, a method independent of transmission distance. These shaped photons enable quantum state transfer with a fidelity of $0.940 \pm 0.008$, and deterministic generation of a Bell state with a fidelity of $0.936 \pm 0.006$. Measurements on this Bell state demonstrate a CHSH correlation of $S = 2.223 \pm 0.023$, exceeding the classical limit by 9.7 standard deviations. The Bell state fidelities for both methods are close to the threshold fidelity of 0.96 for surface code quantum communication \cite{Fowler2010}. This simple yet efficient circuit architecture thus provides a powerful tool to explore complex quantum communication protocols and network designs, and can serve as a testbed for distributed implementations of the surface code.

\begin{figure*}
\begin{center}
	\includegraphics[width=5in]{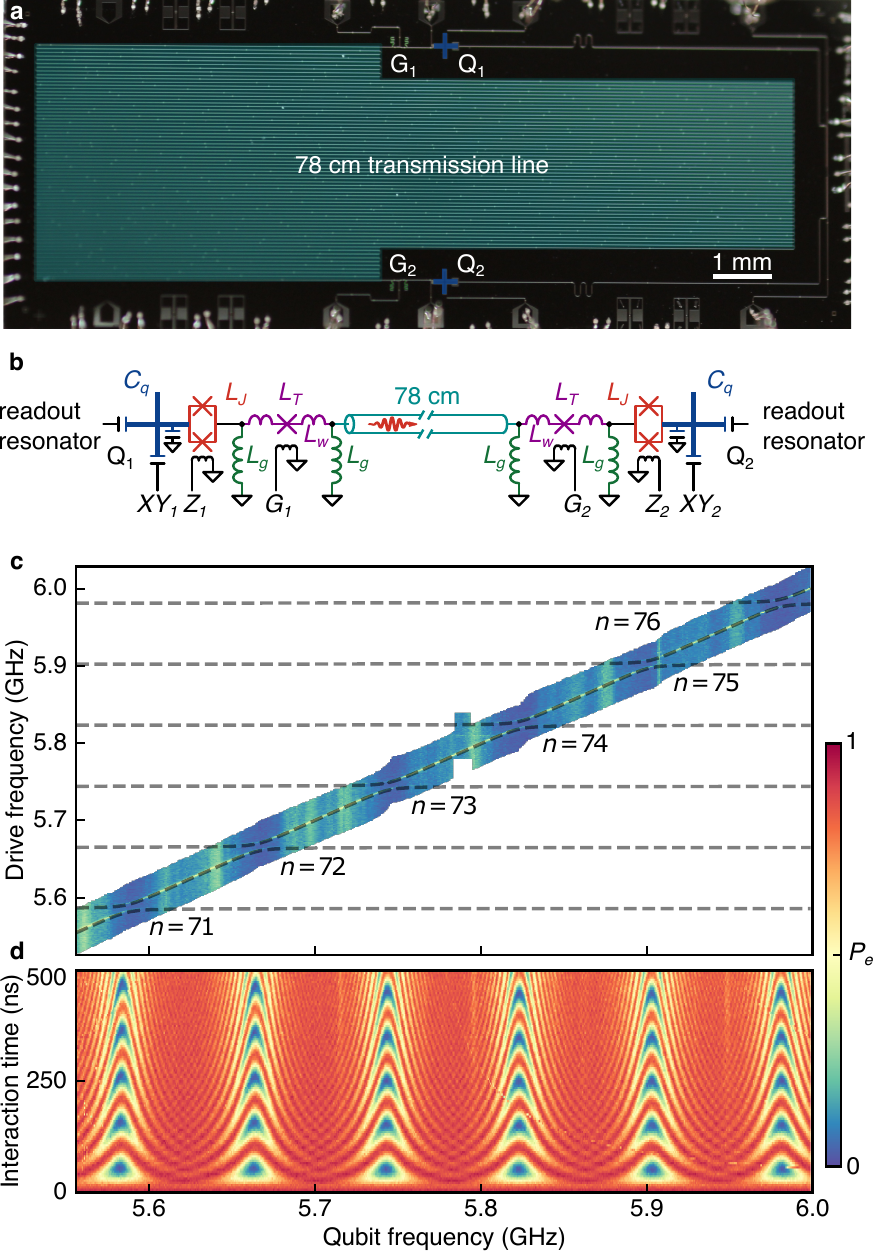}
	\caption{
    \label{fig1}
    {\bf Device description.}
    {\bf a,} Photograph of device, showing two qubits $Q_1$ and $Q_2$ (blue) connected via tunable couplers $G_1$ and $G_2$ (green) to a 78 cm-long coplanar transmission line (cyan).
    {\bf b,} Circuit schematic, with parameters listed in Table~S2 (Supplementary Information).
    {\bf c,} Spectroscopy of qubit $Q_1$ interacting with six transmission line standing modes. Black dashed lines: Numerical simulations.
    {\bf d,} Vacuum Rabi swaps between $Q_1$ and the six standing modes. The coupling is set to $g_1/2\pi = 5$~MHz $\ll \omega_{\rm FSR}/2\pi$.
    }
\end{center}
\end{figure*}

The device layout is shown in Fig.~\ref{fig1}a, comprising two xmon-style qubits\cite{Koch2007,Barends2013}, $Q_1$ and $Q_2$, connected via two tunable couplers \cite{Chen2014}, $G_1$ and $G_2$, to a coplanar waveguide (CPW) transmission line of length $\ell = 0.78$ m. The device is fabricated on a single sapphire substrate, with the serpentine transmission line covering most of the area of a $6 \times 15$~mm$^2$ chip.  A circuit diagram is shown in Fig.~\ref{fig1}b, with more details in the Supplementary Information (SI).

Ignoring the couplers, the transmission line is shorted to ground on both sides, supporting a sequence of standing modes with frequencies $\omega_n/2 \pi$ equally-spaced by $\omega_{\rm FSR}/2\pi = 1/2 T_{\ell} = 79$ MHz, where $T_{\ell} = 6.3$ ns is the photon travel time along the line. The coupling strength $g_{i,n}/2\pi$ between qubit $Q_i$ and the $n$th standing mode is set by external signals to the coupler $G_i$, and further varies with mode number as $\sqrt{n} \propto \sqrt{\omega_n}$. For the experiments here, $n \sim 70$ and the range of $n$ about this value is at most $\sim \pm 5$, so the modes involved in the experiments here all have similar coupling strengths, varying by less than 5\% with $n$; we can therefore represent the coupling to qubit $Q_i$ by a single value $g_i/2\pi$, whose calibrated value ranges from zero to about 47 MHz, as a function of the control signals to $G_i$. More details can be found in the SI.

When one coupler is set to a small non-zero coupling, with $g_i \ll \omega_{\rm FSR}$, and the other coupler is turned off, the coupled qubit can selectively address each standing mode of the transmission line. This is observed by performing qubit spectroscopy, which reveals a sequence of avoided-level crossings with the standing mode resonances (shown for qubit $Q_1$ in Fig.~\ref{fig1}c). In the time domain, we observe vacuum Rabi swaps with each mode by first preparing the qubit in its excited state $|e\rangle$ using a $\pi$ pulse, then setting the qubit frequency by adjusting its $Z$ bias (Fig.~\ref{fig1}d). The weak coupling allows the qubit to interact with each mode separately, with weak interference fringes visible only near frequencies halfway between each mode.

\begin{figure*}[t]
\begin{center}
	\includegraphics[width=5in]{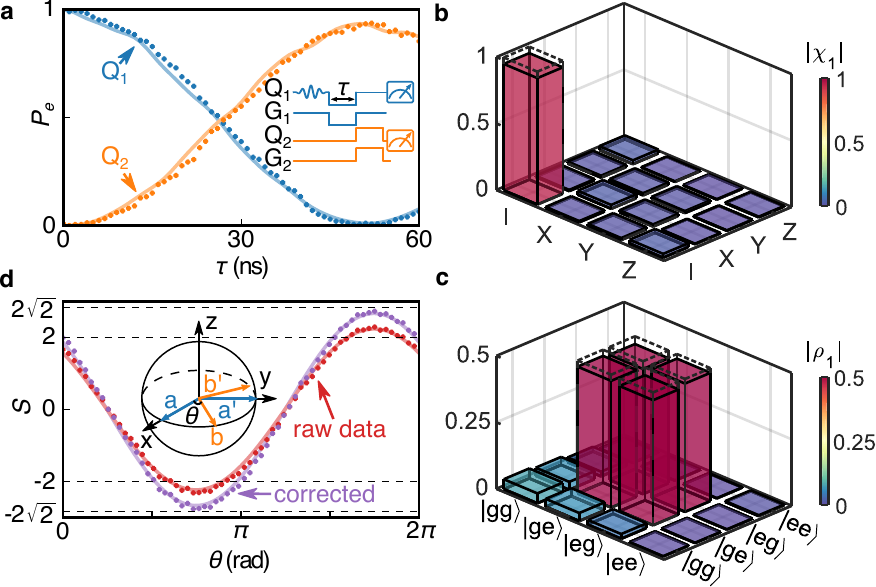}
	\caption{
    \label{fig2}
    {\bf State transfer, remote entanglement, and Bell violation using the relay method.}
    {\bf a,} Quantum state transfer from $Q_1$ to $Q_2$ using the $n = 73, \, \omega_n/2\pi = 5.744$ GHz standing mode as a relay, showing the $|e\rangle$ state probability $P_e$ for each qubit versus swap time $\tau$. Solid lines: Numerical simulations. Inset: Control pulse sequence.
    {\bf b,} Process matrix (absolute values shown as colored bars) of the state transfer process with a fidelity $\mathcal{F}_1^p= 0.952 \pm 0.009$. Dashed-outline frames: Ideal process matrix.
    {\bf c,} Bell state density matrix (absolute values shown as colored bars), with a state fidelity ${\mathcal F}_1^s = 0.950 \pm 0.005$. Dashed-outline frames: Ideal density matrix.
    {\bf d,} Bell test, showing CHSH correlation $S$ versus measurement angle $\theta$. Red dots: No measurement correction; purple dots: With measurement correction. Solid lines: Numerical simulations using $\rho_1$ from panel c. The classical and quantum limits are marked with horizontal dashed lines.
    Inset: Measurement axes $a, a^\prime, b, b^\prime$ on Bloch sphere.
    }
\end{center}
\end{figure*}

By weakly coupling both qubits to a single mode, we can relay qubit states through that mode (Fig.~\ref{fig2}a)~\cite{Sillanpaa2007,Ansmann2009}. We prepare $Q_1$ in its excited state $|e\rangle$, then turn on the $G_1$ coupler for a time $\tau$, while simultaneously adjusting $Q_1$'s frequency to match the selected mode, swapping the excitation to the mode. We then turn on the $G_2$ coupler and adjust $Q_2$'s frequency to swap the excitation to $Q_2$. At $\tau_{\rm swap} = 52$~ns, one photon is completely transferred from $Q_1$ to $Q_2$, with a transfer probability of $0.936 \pm 0.008$. We perform quantum process tomography \cite{Neeley2008} to characterize this transfer process, yielding the process matrix $\chi_1$ shown in Fig.~\ref{fig2}b, with a process fidelity $\mathcal{F}_1^p = \rm{Tr}(\chi_1 \cdot \chi_{\rm{ideal}}) = 0.952 \pm 0.009$. Here $\chi_{\rm ideal}$ is the ideal process matrix for the identity operation $\mathcal{I}$.
Numerical simulations using the master equation give a process fidelity $\mathcal{F}_1^p = 0.955$, in good agreement with experiment (see SI). Note a related experiment \cite{Leung2018} has demonstrated quantum state transfer through a 1~m-long normal-metal coaxial cable using a hybridized ``dark'' relay mode, achieving a transfer fidelity of 0.61 with a significantly lossier channel.

We also use the relay mode to generate a Bell singlet state $|\psi_{\rm Bell}^{-}\rangle = \left (|ge\rangle-|eg\rangle \right )/\sqrt{2}$ between the two qubits, by terminating the $Q_1$ swap process at the half-swap time $\tau_{\rm half} = 26$~ns. We perform quantum state tomography \cite{Steffen2006b}, with the reconstructed density matrix $\rho_1$ displayed in Fig.~\ref{fig2}c, from which we calculate a state fidelity $\mathcal{F}_1^s = \langle \psi_{\rm Bell}^{-} |\rho_1|\psi_{\rm Bell}^{-}\rangle = 0.950 \pm 0.005$ and a concurrence ${\mathcal C}_1 = 0.927 \pm 0.013$.  This experimental result agrees well with the numerically-simulated state fidelity $\mathcal{F}_1^s = 0.947$ and concurrence ${\mathcal C}_1 = 0.914$.

We next perform the CHSH Bell inequality test \cite{Ansmann2009} on this remotely entangled Bell state (see SI). We measure $Q_1$ along direction $a = x$ or $a^\prime = y$, and simultaneously measure $Q_2$ along $b$ or $b^\prime \bot b$, varying the angle $\theta$ between $a$ and $b$ (Fig. 2D inset). We then calculate the CHSH correlation $S$, as shown in Fig.~\ref{fig2}d. We find that $S$ is maximized at $\theta=5.5$~rad, very close to the ideal value of $7\pi/4 \cong 5.498$, where $S = 2.237 \pm 0.036$ with no measurement correction, exceeding the maximum classical value of 2 by 6.6 standard deviations. If we correct for readout error \cite{Ansmann2009}, we find $S = 2.665 \pm 0.044$, approaching the quantum limit of $2\sqrt{2}\cong 2.828$. The entanglement is deterministic and the measurement is single-shot (see SI), so the detection loophole \cite{Garg1987} is closed in this experiment.

The relay method requires $g_i \ll \omega_{\rm FSR}$ so that the swap process only involves a single mode. However, $\omega_{\rm FSR}$ scales inversely with transmission distance $\ell$, making $g_i$ impractically small as $\ell$ increases. An alternative approach, independent of transmission distance, is to use itinerant photons for state transfer \cite{Cirac1997,Korotkov2011}. This is experimentally challenging, and has only recently been demonstrated with superconducting qubits \cite{Kurpiers2018, Axline2018, Campagne2018}. In these experiments, quantum states were transferred through a $\sim 1$~m-long superconducting coaxial cable interrupted by a circulator. The state transfer speeds were significantly slower than the photon travel time in the channel, making reflections and their interference nearly unmanageable without the circulator. The circulator however also introduces loss, limiting transfer fidelities to about 80\%. With the one to two orders of magnitude stronger coupling achieved here, enabling photon transfers in less than the photon round-trip travel time, we perform remote state transfer and entanglement generation using shaped itinerant photons without a circulator, achieving sufficient fidelity to violate the Bell inequality.

\begin{figure*}[t]
\begin{center}
	\includegraphics[width=5in]{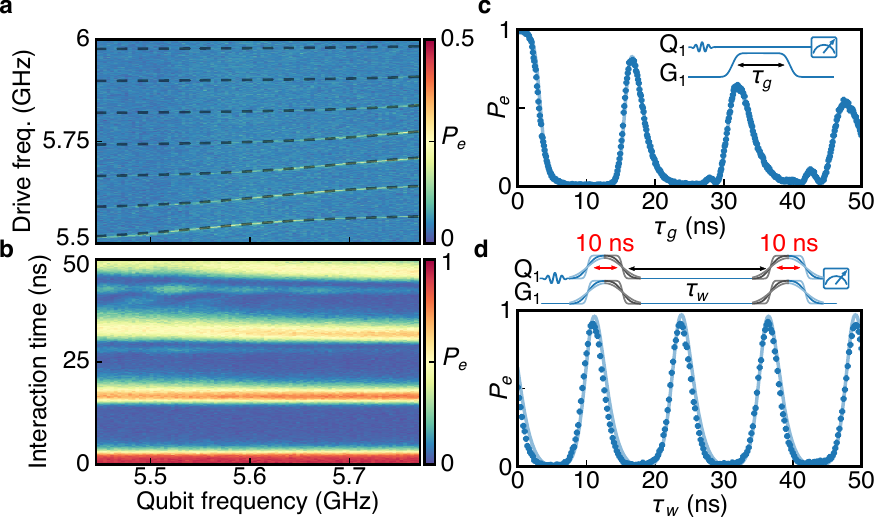}
	\caption{
    \label{fig3}
    {\bf Single qubit ``ping-pong'' with itinerant photons.}
    {\bf a,} Qubit $Q_1$ spectrum when strongly coupled to the transmission line, showing multiple modes interacting with the qubit. Dashed lines: Numerical simulations.
    {\bf b,} Quantum time-domain reflectometry of $Q_1$ with the transmission line. The coupling is sufficiently strong that the interaction is essentially independent of the qubit frequency.
    {\bf c,} Variant of the reflectometry ``ping-pong'' dynamics, with $Q_1$'s frequency initially set to $5.809$ GHz with the qubit in $|e\rangle$, following which its $Z$ bias remains unchanged. $Q_1$ emits an itinerant photon in about 8~ns, which is reflected from the far end of the transmission line and caught by $Q_1$ a time $2 T_{\ell} = 12.6$ ns later, the process here repeated three times. Solid line: Numerical simulations. Inset: Control pulse sequence, with the rise and fall times indicated; the qubit frequency is changed by the coupling control signals.
    {\bf d,} Optimizing photon catch by adjusting control pulse envelope. Maximum catch probability is improved from $\sim 0.8$ in panel {\bf c} to $0.922\pm 0.004$ by adjusting the control pulse slope. Solid line: Numerical simulations. Top: Control pulse sequence, showing the pulse shaping. The rising edge of the first control pulse determines the emission process, and the falling edge of the second control pulse determines the capture process. The qubit bias pulses cancel the coupler-generated frequency shift (see SI).}
\end{center}
\end{figure*}

In Fig. 3 we show the first part of the itinerant photon method, tuning $Q_1$'s interaction with the transmission line so that $Q_1$ can play single-photon ``ping-pong'' with itself. In Fig.~\ref{fig3}a we show the qubit-transmission line spectroscopy, measured at maximum coupling $|g_1|/2\pi = 47$~MHz, with $Q_2$'s coupler turned off. In this regime, the avoided-level crossing with each mode (Fig.~\ref{fig1}c) disappears; instead multiple modes are coupled with the qubit. In Fig.~\ref{fig3}b we perform quantum time-domain reflectometry, where $Q_1$ is excited to $|e\rangle$, then we immediately turn $G_1$'s coupling to its maximum value while fixing the qubit frequency by adjusting the qubit $Z$ bias, both for a duration $\tau_g$, following which we monitor the qubit response. The qubit excitation is released into the transmission line in a few nanoseconds, leaving the qubit in its ground state $|g\rangle$ until the photon reflects off the far end of the transmission line and returns to the qubit, re-exciting the qubit to its $|e\rangle$ state. This process does not depend on qubit frequency, other than some small features.

In Fig.~\ref{fig3}c, we perform a variant of the reflectometry ``ping-pong'' experiment, where after exciting $Q_1$ to $|e\rangle$, we leave $Q_1$'s $Z$ bias fixed, allowing $Q_1$'s frequency to vary due to changes in the coupling; see SI. We see that the emission takes about 8 ns, with the round trip then completed in $2 T_{\ell} = 12.6$~ns.  Three full transits are shown, with the peak amplitude falling and small ripples appearing, mainly due to scattering from each photon-qubit interaction. The coupling here is strong enough that the rise and fall time of the control pulse must be accounted for in the simulations (see SI).

Next, to tune up the photon emission and capture process, we set the emission and capture times to 10 ns, and vary the wait time $\tau_w$ between them. We dynamically tune $Q_1$'s coupling, while keeping the qubit frequency fixed (see SI). Ideally, the itinerant photon can be captured with unit probability if the emission and capture control pulses are properly tuned \cite{Cirac1997,Korotkov2011}. However, the bandwidth of our control electronics is insufficient to allow the desired sub-nanosecond tuning of the itinerant photon envelope, so we instead approximately tune the coupling by convolving a Gaussian and a rectangle pulse; the width of the Gaussian shapes the edges of the convolved pulse (see SI). We find that this sub-optimal shaping still achieves a self-capture probability of $0.922 \pm 0.004$ (Fig.~\ref{fig3}d). The robustness of this protocol to control pulse imperfections is as expected~\cite{Sete2015}.

\begin{figure*}[t]
\begin{center}
	\includegraphics[width=5in]{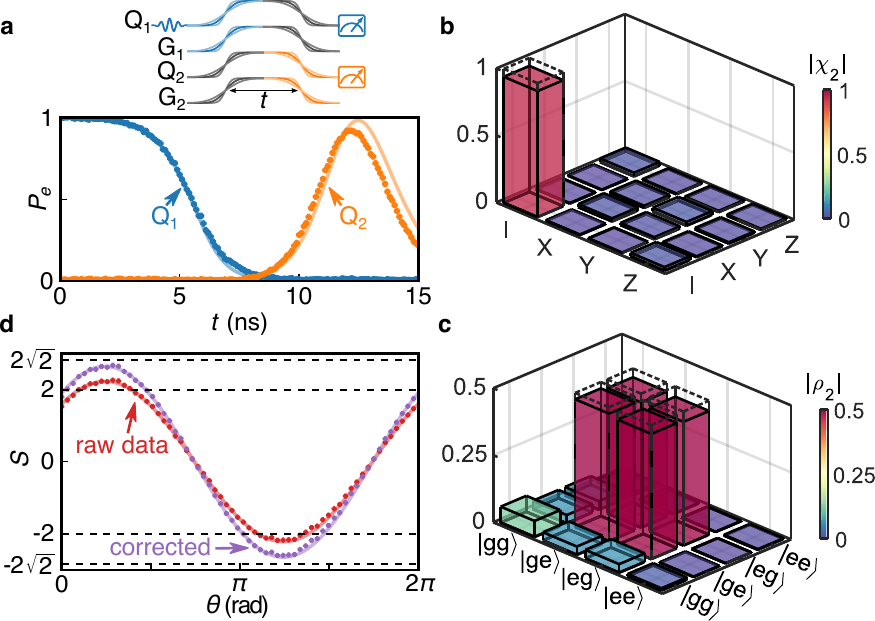}
	\caption{
    \label{fig4}
    {\bf State transfer, remote entanglement and Bell violation using itinerant photons.}
    {\bf a,} Using optimized control pulses for the couplers while keeping the qubit frequency fixed, we achieve a high-fidelity state transfer, with a maximum transfer probability of $0.919 \pm 0.004$ in 12.2 ns. Solid lines: Numerical simulations. Top: Control pulse sequence. The rising edge of $Q_1$'s control pulse determines the emission process, and the falling edge of $Q_2$'s control pulse determines the capture process.
    {\bf b,} Quantum process tomography, with a process fidelity ${\mathcal F}_2^p = 0.940 \pm 0.008$. Dashed-outline frames: Ideal process matrix.
    {\bf c,} Density matrix of the Bell state generated by sending half an itinerant photon from $Q_1$ to $Q_2$, with a state fidelity ${\mathcal F}_2^s = 0.936 \pm 0.006$ and a concurrence $\mathcal{C}_2 = 0.914 \pm 0.014$. Dashed-outline frames: Ideal density matrix.
    {\bf d,} Bell test, showing the CHSH correlation $S$ versus measurement angle $\theta$. Red dots: No measurement correction; purple dots: With measurement correction. Solid lines: Numerical simulations using $\rho_2$ from panel {\bf c}. The correlation is maximized at $\theta=0.84$ rad, where $S = 2.223 \pm 0.023$ without measurement correction. Classical and quantum limits are marked with horizontal dashed lines.
    }
\end{center}
\end{figure*}

We perform this tune-up for each qubit separately, then combine these processes to perform qubit-to-qubit state transfer using itinerant photons (Fig.~\ref{fig4}a). We first excite $Q_1$ to $|e\rangle$, with $Q_2$ in $|g\rangle$, then turn on the tuned $G_1$ and $G_2$ time-dependent couplings simultaneously for a duration $t$. The itinerant photon is released from $Q_1$ into the channel in about 10~ns, and begins to interact with $Q_2$ after $T_{\ell} = 6.3$ ns. The photon is captured by $Q_2$, with a maximum probability of $0.919 \pm 0.004$ at $t=12.2$~ns. We carry out quantum process tomography for this sequence, and reconstruct the process matrix $\chi_2$ (see Fig.~\ref{fig4}b), with a fidelity $\mathcal{F}_2^p = \rm{Tr}(\chi_2 \cdot \chi_{\rm{ideal}}) = 0.940 \pm 0.008$. Finally, we use half an itinerant photon to generate entanglement between the two qubits: We first prepare $Q_1$ in $|e\rangle$, then control $Q_1$'s coupling to release half its excitation to the channel, which is captured by $Q_2$ using the same time-domain coupling as in the state transfer experiment. This generates a Bell triplet state $|\psi_{\rm Bell}^{+}\rangle=(|ge\rangle+|eg\rangle)/\sqrt{2}$ between the two qubits (Fig.~\ref{fig4}c), with a reconstructed Bell state fidelity $\mathcal{F}_2^s = \langle \psi_{\rm Bell}^{+} |\rho_2|\psi_{\rm Bell}^{+}\rangle = 0.936 \pm 0.006$ and a concurrence $\mathcal{C}_2 = 0.914 \pm 0.014$.

As with the relay mode method, we carry out a CHSH Bell inequality test with no detection loophole \cite{Garg1987}. We find that $S$ is maximized at $\theta=0.84$ rad, close to the ideal value of $\pi/4 \cong 0.785$, where $S = 2.223 \pm 0.023$ without applying a measurement correction, exceeding the classical limit of 2 by 9.7 standard deviations. If we correct for readout error, we find $S = 2.629 \pm 0.028$, close to the quantum limit of $2\sqrt{2}$.

In conclusion, we present a simple architecture that allows efficient quantum state transfer and remote entanglement between two superconducting qubits, connected by a 78 cm-long transmission line. The fidelities are sufficient to violate the Bell inequality using two different methods. This architecture can be expanded to multiple communication channels, allowing the exploration of more complex quantum communication protocols, and could serve as a backbone for fault-tolerant distributed quantum computing.

\clearpage

\bibliography{bibliography}
\bibliographystyle{naturemag}

\clearpage

\subsection*{Acknowledgements}
We thank P. J. Duda for fabrication assistance, and Y. Lu and S. Chakram for helpful discussions. We thank MIT Lincoln Laboratory for providing a traveling-wave parametric amplifier. This effort is supported by the Army Research Office under contract W911NF-15-2-0058. The views and conclusions contained in this document are those of the authors and should not be interpreted as representing the official policies, either expressed or implied, of the Army Research Laboratory or the U.S. Government. The U.S. Government is authorized to reproduce and distribute reprints for Government purposes notwithstanding any copyright notation herein. Devices and experiments were also supported by the Air Force Office of Scientific Research, and by the Department of Energy (DOE). K.J.S. was supported by NSF GRFP (NSF DGE-1144085), \'E.D. was supported by LDRD funds from Argonne National Laboratory, A.N.C. was supported in part by the DOE, Office of Basic Energy Sciences, and D.I.S. acknowledges support from the David and Lucile Packard Foundation. This work was partially supported by the UChicago MRSEC (NSF DMR-1420709) and made use of the Pritzker Nanofabrication Facility, which receives support from SHyNE, a node of the National Science Foundation's National Nanotechnology Coordinated Infrastructure (NSF NNCI-1542205).

\subsection*{Author Contributions}
Y.P.Z. designed and fabricated the devices. Y.P.Z, H.S.C., K.J.S., M.H.C., J.G. and A.N.C. developed the fabrication processes. H.S.C., K.J.S. and A.N.C. contributed to device design. Y.P.Z. performed the experiments and analyzed the data. A.N.C. and D.I.S. advised on all efforts. All authors contributed to discussions and production of the manuscript.

\subsection*{Author Information}
The authors declare no competing financial interests. Correspondence and requests for materials should be addressed to A. N. Cleland (anc@uchicago.edu).
The datasets supporting this work are available from the corresponding author upon request.

\end{document}



\title{Supplementary Information for ``Violating Bell's inequality with remotely-connected superconducting qubits''}

\author{Y. P. Zhong}
\affiliation{Institute for Molecular Engineering, University of Chicago, Chicago IL 60637, USA}
\author{H.-S. Chang}
\affiliation{Institute for Molecular Engineering, University of Chicago, Chicago IL 60637, USA}
\author{K. J. Satzinger}
\affiliation{Institute for Molecular Engineering, University of Chicago, Chicago IL 60637, USA}
\affiliation{Department of Physics, University of California, Santa Barbara, CA 93106, USA}
\author{M.-H. Chou}
\affiliation{Institute for Molecular Engineering, University of Chicago, Chicago IL 60637, USA}
\affiliation{Department of Physics, University of Chicago, Chicago IL 60637, USA}
\author{A. Bienfait}
\affiliation{Institute for Molecular Engineering, University of Chicago, Chicago IL 60637, USA}
\author{C. R. Conner}
\affiliation{Institute for Molecular Engineering, University of Chicago, Chicago IL 60637, USA}
\author{\'E. Dumur}
\affiliation{Institute for Molecular Engineering, University of Chicago, Chicago IL 60637, USA}
\affiliation{Institute for Molecular Engineering and Materials Science Division, Argonne National Laboratory, Argonne IL 60439, USA}
\author{J. Grebel}
\affiliation{Institute for Molecular Engineering, University of Chicago, Chicago IL 60637, USA}
\author{G. A. Peairs}
\affiliation{Institute for Molecular Engineering, University of Chicago, Chicago IL 60637, USA}
\affiliation{Department of Physics, University of California, Santa Barbara, CA 93106, USA}
\author{R. G. Povey}
\affiliation{Institute for Molecular Engineering, University of Chicago, Chicago IL 60637, USA}
\affiliation{Department of Physics, University of Chicago, Chicago IL 60637, USA}
\author{D. I. Schuster}
\affiliation{Department of Physics, University of Chicago, Chicago IL 60637, USA}
\author{A. N. Cleland}
\affiliation{Institute for Molecular Engineering, University of Chicago, Chicago IL 60637, USA}
\affiliation{Institute for Molecular Engineering and Materials Science Division, Argonne National Laboratory, Argonne IL 60439, USA}

\maketitle

\setcounter{equation}{0}
\setcounter{figure}{0}
\setcounter{table}{0}
\setcounter{page}{1}

\renewcommand{\theequation}{S\arabic{equation}}
\renewcommand{\thefigure}{S\arabic{figure}}
\renewcommand{\thetable}{S\arabic{table}}

\section{Comparison with similar experiments}
There have been a number of recent experiments demonstrating deterministic remote state transfer and entanglement generation with superconducting qubits. In Table~\ref{tableone} we tabulate the main results of these experiments and compare with the results reported here.

\begin{table}[H]
\begin{center}
\begin{tabular}{|l|c|c|c|c|c|c|}
  \hline
  \hline
  Source &coupling rate & Transfer   & Process                   & State                     & Concurrence    & CHSH \\
         &$\kappa/2\pi$ ($g/2\pi$)& efficiency & fidelity ${\mathcal F}^p$ & fidelity ${\mathcal F}^s$ & ${\mathcal C}$ &  correlation $S$  \\
  \hline
  \hline
  This paper   &(5 MHz)& 0.936 & 0.952 & 0.950 & 0.927 & 2.237 \\
  (relay mode) &     &      &       &       &       & \\
  \hline
  This paper        & $\sim 175$ MHz & 0.919 & 0.940 & 0.936 & 0.914 & 2.223 \\
  (itinerant photon)&       &       &       &       &       & \\
  \hline
  Kurpiers {\it et al.} \cite{Kurpiers2018} & $\sim 10$ MHz & 0.676 & 0.8002 & 0.789 & 0.747 & N/A \\
  \hline
  Axline {\it et al.} \cite{Axline2018}     &$\sim 1$ MHz   & 0.74 & 0.76 & 0.61 & 0.51 & N/A \\
  \hline
  Campagne-Ibarcq {\it et al.} \cite{Campagne2018} &$\sim 1$ MHz& 0.7 & N/A & 0.73 & N/A & N/A \\
  \hline
  Leung {\it et al.} \cite{Leung2018} & ($\sim 2$ MHz) & N/A & 0.61 & 0.793 & N/A & N/A \\
  \hline
  \hline
\end{tabular}
\end{center}
\caption{\label{tableone} Comparison of similar deterministic remote state transfer and entanglement generation experiments on superconducting circuits. Here $\kappa/2\pi$ is the photon decay rate into the channel (itinerant photon method), $g/2\pi$ is the on-resonant coupling between the qubit and the relay mode, ${\mathcal F}^p$ is the state transfer process fidelity, ${\mathcal F}^s$ the Bell state fidelity, ${\mathcal C}$ the Bell state concurrence, and $S$ the CHSH correlation.}
\end{table}

\section{Device fabrication}
Most of the fabrication is done on 100~mm-diameter sapphire substrates, with steps 5-7 typically completed on quarters cut from the larger wafer. This recipe is adapted in part from Refs. \onlinecite{Kelly2015thesis, Dunsworth2018}.
\begin{enumerate}
\item $100$~nm Al base layer deposition using electron beam evaporation.
\item Base layer lithography and dry etch with BCl$_3$/Cl$_2$/Ar inductively coupled plasma. This defines the qubit capacitors, the tunable coupler wiring, the 78 cm-long transmission line, and the readout and control circuitry.
\item $1~\mu$m crossover scaffold SiO$_2$ deposition using electron beam evaporation and liftoff, using an optically-patterned PMMA/nLOF2020 bilayer. The thin PMMA layer serves as a protection layer for the base Al layer from step 1 during the development of nLOF2020 in AZ300 MIF. The PMMA is then removed with a downstream O$_2$ plasma ash after development.
\item $500$~nm crossover Al deposition with the same liftoff patterning method as step 3. The Al deposition is preceded by an {\it in situ} Ar-ion mill without breaking vacuum between these two steps.
\item Josephson junction deposition using the Dolan bridge method \cite{dolan1977} shadow evaporation and liftoff, using a PMMA/MAA bilayer and electron beam lithography. The Al evaporated in this step does not contact the base wiring and is not preceded by an Ar ion mill.
\item Bandage Al liftoff deposition\cite{Dunsworth2017}, preceded by an {\it in situ} Ar ion mill. This step establishes galvanic connections between the base wiring Al from step 1 and the Josephson junctions defined in step 5.
\item Vapor HF to remove the SiO$_2$ scaffold underlying the Al crossovers.
\end{enumerate}

We use electron beam evaporation to deposit each film. We use photolithography with $0.9~\mu$m I-line photoresist (AZ MiR 703) for steps 2 and 6. Each liftoff step is in N-methyl-2-pyrrolidone at $80^\circ$C.

\section{Experimental setup}
\begin{figure}[H]
  \begin{center}
  \includegraphics[width=4.5in]{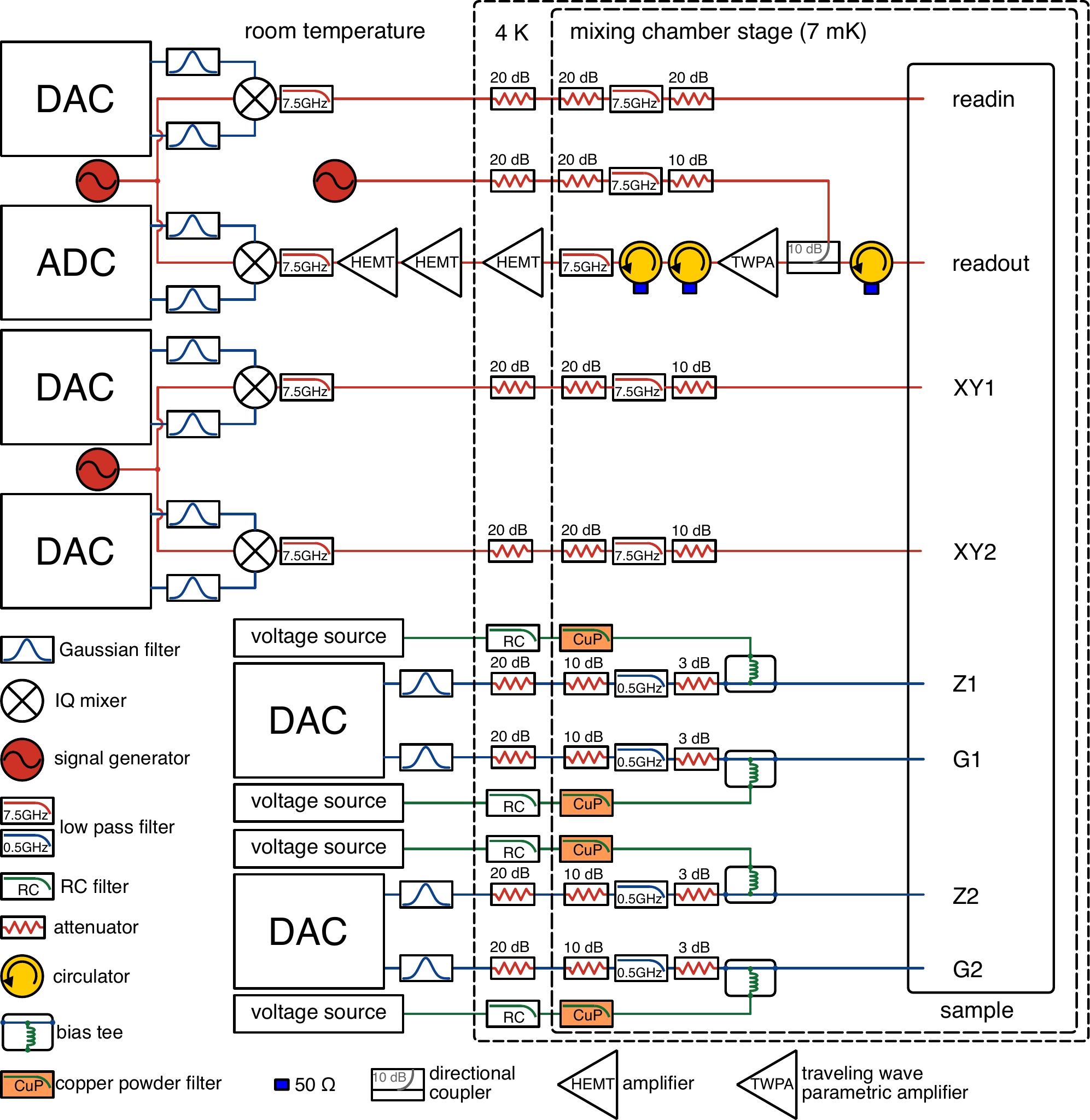}
  \caption{\label{figs1} Electronics and wiring. Red lines correspond to radiofrequency (RF) and microwave signals for qubit $XY$ control and measurement, blue lines correspond to intermediate frequency (IF) signals for fast qubit $Z$ or coupler control, and green lines correspond to quasi-DC signals for steady qubit $Z$ or coupler bias offset. The IF and DC signals for each bias channel are combined using a custom-made cryogenic bias tee mounted at the mixing chamber stage.}
  \end{center}
\end{figure}

Figure~\ref{figs1} shows the overall control and readout electronics layout. We use custom digital-to-analog converter (DAC) and analog-to-digital converter (ADC) circuit boards for qubit control and measurement, respectively. The control boards have dual-channel 14-bit vertical resolution DAC integrated circuits operating at 1~Gs/s, and the measurement boards have dual-channel 8-bit vertical resolution ADC integrated circuits operating at 1~Gs/s. Each control signal output and measurement signal input channel is filtered by a custom Gaussian low-pass filter with 250 MHz bandwidth. The control boards are used to generate nanosecond-length pulses for fast qubit $Z$ or coupler $G$ control, or to provide the modulation envelope for several-GHz carrier signals, the two combined using an IQ mixer.  In this application the signals are used to implement qubit $XY$ rotations, or to drive the readout resonator feed-line for qubit state measurements. In the latter case, the output signal from the readout feed-line is first amplified by a traveling wave parametric amplifier \cite{macklin2015} (TWPA) at the mixing chamber stage with close to quantum-limited added noise, then amplified by a cryogenic high electron mobility transistor (Low Noise Factory HEMT) at the 4~K stage, and further amplified by two room-temperature Miteq HEMT amplifiers, before down-conversion with an IQ mixer and capture by the measurement ADC board. Two cryogenic circulators with low insertion loss are added between the TWPA and the cryogenic HEMT to block reflections as well as noise emitted from the input of the cryogenic HEMT. An additional circulator is inserted between the TWPA drive line and the qubit, to avoid any unexpected excitation of the qubits from the TWPA drive signal. The measurement board has an on-board demodulation function which allows for fast demodulation of the captured waveform. Each control line is heavily attenuated and filtered at each temperature stage in the dilution refrigerator to minimize the impact on the qubit coherence while retaining controllability.

\section{Device Characterization}
Each qubit can be tuned from 3 to 7.3 GHz using its $Z$-control current bias, with full quantum state control using the $XY$-control microwave drive line, and dispersive readout with a capacitively-coupled readout resonator \cite{Koch2007}.

\subsection{Summary of device parameters}
In Table \ref{parameters} we display the characteristics for each qubit $Q_1$ and $Q_2$. Parameters preceded by $^*$ are design values; others are experimentally determined.

\begin{table}[H]
\begin{center}
\begin{tabular}{l c c c c c}
  \hline
  \hline
  Parameters & $Q_1$ & $Q_2$ \\
  \hline
  $^\ast$Qubit capacitance, $C_q$ & 90 fF & 90 fF\\
  Qubit junction inductance, $L_J$ & 8.34 nH & 8.57 nH \\
  $^\ast$Coupler inductance to ground, $L_g$ & 0.2 nH & 0.2 nH \\
  $^\ast$Coupler stray wiring inductance, $L_w$ & 0.1 nH & 0.1 nH \\
  Coupler junction inductance, $L_T$  & 0.566 nH  & 0.564 nH \\
  Qubit operating frequency, $\omega_i/2\pi$ & 5.809 GHz & 5.731 GHz\\
  Qubit anharmonicity, $\alpha$ & -160 MHz & -162 MHz\\
  Qubit lifetime, $T_1$ & 16 $\mu$s & 11 $\mu$s\\
  Qubit Ramsey dephasing time, $T_2$ & 0.89 $\mu$s & 0.85 $\mu$s\\
  Readout resonator frequency, $\omega_r/2\pi$ & 6.4527 GHz & 6.3390 GHz\\
  $^\ast$Readout coupling, $g_r/2\pi$ & 38 MHz & 38 MHz \\
  Readout dispersive shift, $\kappa_r$ & 0.6 MHz & 0.8 MHz\\
  $|g\rangle$ state readout fidelity, $F_g$ & 0.984 & 0.984 \\
  $|e\rangle$ state readout fidelity, $F_e$ & 0.950 & 0.942 \\
  \hline
  $^\ast$ These are design parameters.
\end{tabular}
\end{center}
\caption{\label{parameters} Device parameters.}
\end{table}

\subsection{Qubit single-shot readout}
\begin{figure}
  \begin{center}
  \includegraphics[width=4.5in]{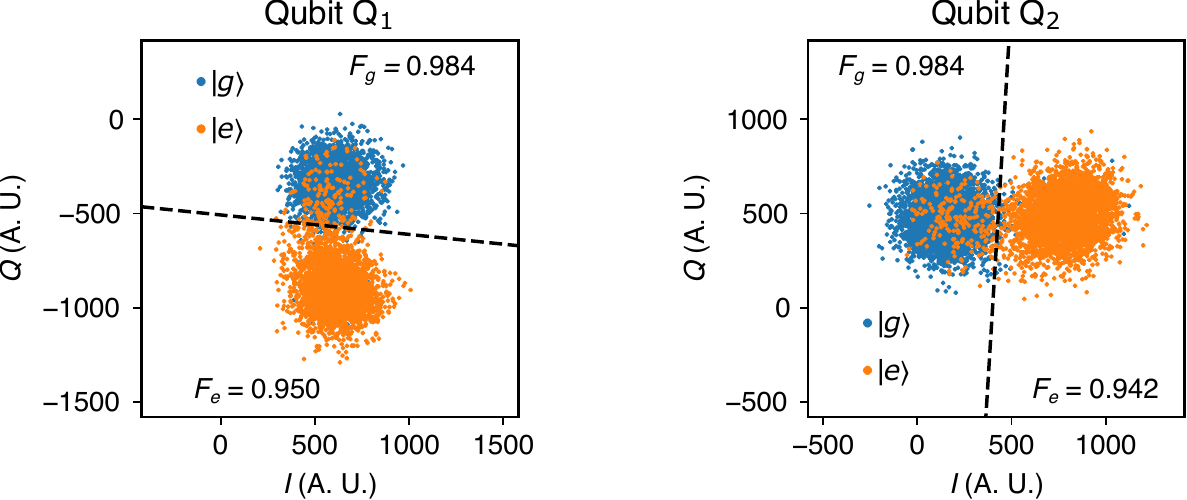}
  \caption{\label{figs2} Qubit single-shot readout. A large number of measurements were made with each qubit in either its ground $|g\rangle$ or its excited $|e\rangle$ state, and data accumulated as the blue or orange points respectively. This calibration allows us to assign any given measurement to the ground or excited state, as separated by the dashed lines in the $IQ$ plane. For $Q_1$, the $|g\rangle$ state readout fidelity is $F_g=0.984$, and the $|e\rangle$ state readout fidelity is $F_e=0.950$. For $Q_2$, the $|g\rangle$ state readout fidelity is $F_g=0.984$, and the $|e\rangle$ state readout fidelity is $F_e=0.942$.}
  \end{center}
\end{figure}

We characterize the qubit readout fidelity by turning the coupler for each qubit as close to zero as possible, to isolate the qubit from the rest of the circuit. With the qubit in its equilibrium state (mostly in its ground state $|g\rangle$), we then perform a standard single-shot readout measurement, and record the values of the microwave quadratures $I$ and $Q$ corresponding to the readout result. We accumulate a large number of these events, shown in blue in Fig.~\ref{figs2}. We then repeat this process, but precede the measurement with an on-resonant microwave pulse calibrated to put the qubit in its excited state $|e\rangle$. The results of these measurements are shown in orange in Fig.~\ref{figs2}. These calibrations allow us to assign any single-shot measurement, based on its de-convolved $I$ and $Q$ values, to the $|g\rangle$ or $|e\rangle$ state based on which side of the dashed line in Fig.~\ref{figs2} the measurement falls. For $Q_1$, the $|g\rangle$ state readout fidelity is $F_g=0.984$, and the $|e\rangle$ state readout fidelity is $F_e=0.950$. For $Q_2$, the $|g\rangle$ state readout fidelity is $F_g=0.984$, and the $|e\rangle$ state readout fidelity is $F_e=0.942$.

\subsection{Multimode transmission line}
The 78~cm-long coplanar waveguide transmission line used in this experiment has a 4~$\mu$m-wide center trace and a 2~$\mu$m gap to the ground plane on each side, with specific capacitance $\mathscr{C} = 173$~pF/m and specific inductance $\mathscr{L} = 402$~nH/m. Neglecting the coupler, the line is shorted by $L_g$ at its far end, where this inductance is provided by a short segment of transmission line. We absorb this length in the overall transmission line, so that the input impedance is given by
\begin{equation}\label{input_impedance}
  Z_{\rm in} = Z_0\tanh(\alpha + i \beta) \ell = Z_0\frac{\tanh\alpha \ell + i \tan \beta \ell}{1+i \tan (\beta \ell) \tanh (\alpha \ell)},
\end{equation}
where $\alpha + i \beta$ is the complex propagation parameter, and $Z_0=\sqrt{\mathscr{L}/\mathscr{C}}$ is the characteristic impedance of the transmission line \cite{Pozar}.

\begin{figure}
  \begin{center}
  \includegraphics[width=4.5in]{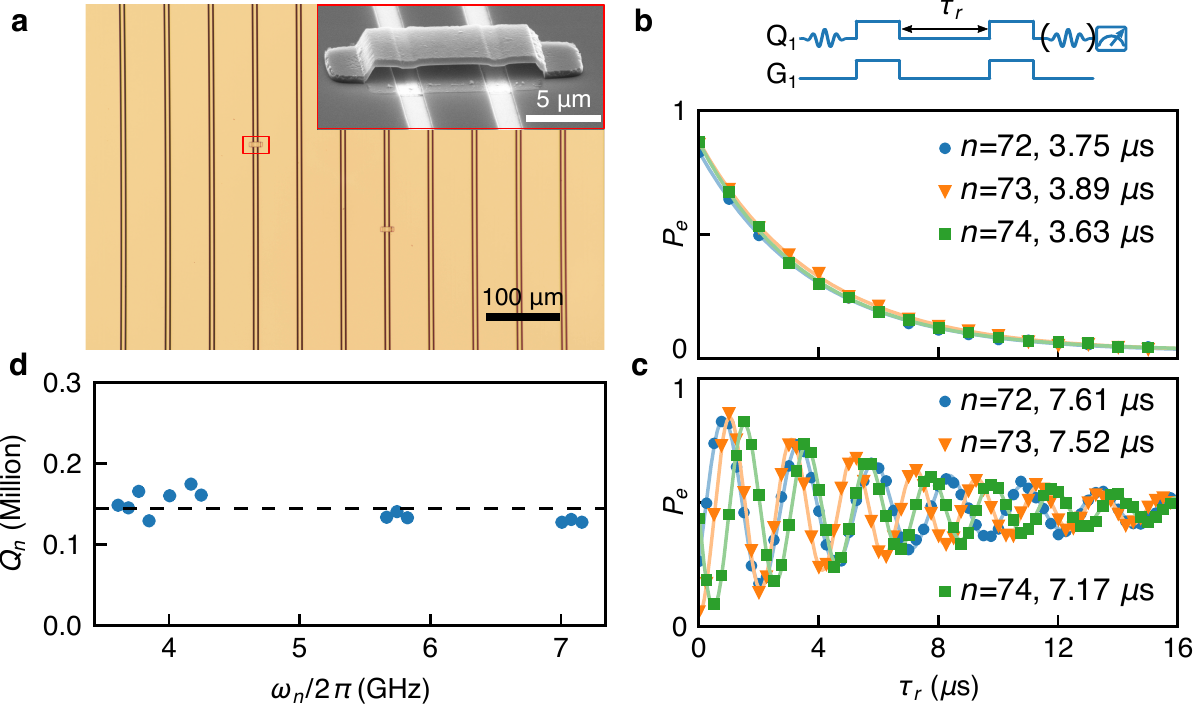}
  \caption{\label{figs3} Transmission line characterization.
  {\bf a,} Optical micrograph of a small portion of the transmission line, which has a 4~$\mu$m wide center trace and a 2~$\mu$m gap to the ground plane on either side. The transmission line meanders are separated by $60~\mu$m, and the line has 390 air-bridge crossovers evenly distributed along the line every 2 mm, suppressing unwanted slot-line modes and other microwave resonances. Inset: Scanning electron micrograph picture of an air-bridge crossover.
  {\bf b,}, {\bf c,} The lifetime $T_{1n}$ and Ramsey dephasing time $T_{2n}$ of three of the six resonant modes shown in Fig.~1d. We find $T_{2n} \approx 2T_{1n}$, indicating negligible dephasing noise in the transmission line. Solid lines: Fits to each mode's data. Top: Control pulse sequence.
  {\bf d,} Quality factor $Q_n=\omega_n T_{1n}$ measured for different modes from 3.6 GHz to 7.2 GHz. We find that the quality factor is more or less constant over this frequency range, with an average $\langle Q \rangle \sim 1.44\times 10^5$ as indicated by the horizontal dashed line.
  }
  \end{center}
\end{figure}

Near the $n$th mode resonance,
\begin{equation}\label{nmode}
  \beta \ell = n\pi + \frac{\pi\Delta \omega}{\omega_{\lambda/2}},
\end{equation}
where $\omega_{\lambda/2}$ is the half-wave radial frequency. Near this frequency we have the input impedance
\begin{equation}\label{Zinapprox}
  Z_{\rm in} \approx Z_0 \left (\alpha \ell + i \frac{\pi\Delta \omega}{\omega_{\lambda/2}} \right ),
\end{equation}
where we assume $\alpha \ell \ll 1$, a safe assumption for a superconducting transmission line on a very low-loss substrate such as sapphire.

This impedance is equivalent to a series $RLC$ resonant circuit with equivalent lumped-element parameters
\begin{eqnarray}
  \omega_n &=& n \omega_{\lambda/2}, \\
  R_n &=& Z_0\alpha \ell, \\
  L_n &=& \frac{\pi Z_0}{2\omega_{\lambda/2}}=\frac{1}{2} \mathscr{L} \ell, \\
  C_n &=& \frac{1}{n^2\omega_{\lambda/2}^2L_n}, \\
  Q_n &=& \frac{\omega_n L_n}{R_n} = \frac{\beta}{2\alpha}. \label{Qeq}
\end{eqnarray}

In Fig.~\ref{figs3}, we display the transmission line and its characterization. Figure~\ref{figs3}a shows an optical micrograph of a small portion of the transmission line and a scanning electron micrograph picture of one of the 390 air-bridge crossovers evenly distributed along the line. In Fig.~\ref{figs3}b and c,we use $Q_1$, weakly coupled to the line, to measure the lifetime $T_{1n}$ and the Ramsey dephasing time $T_{2n}$ of three resonator modes, with $T_{2n}\approx 2T_{1n}$ indicating that dephasing noise is negligible in the channel. In Fig.~\ref{figs3}d, we show the quality factor $Q_n = \omega_{n}T_{1n}$ for different modes ranging from $3.6$ GHz to $7.2$ GHz. We find that $Q_n$ is more or less constant over this span of frequencies, with an average $\langle Q \rangle \sim 1.44\times 10^5$. Comparing to Eq.~(\ref{Qeq}), this suggests that the attenuation parameter $\alpha$ has a linear frequency dependence similar to $\beta$, indicating that dielectric loss dominates in this frequency range \cite{Pozar}. We note that similar quality factors can be achieved with superconducting coaxial cables \cite{Kurpiers2017}, so in principle the transmission line here can be replaced by a superconducting cable for inter-chip quantum communication. Note that for planar transmission lines that include crossovers where the SiO$_2$ dielectric is left as a support structure, measurements find quality factors roughly one order of magnitude smaller than here \cite{Dunsworth2018}. Removing the SiO$_2$ crossover scaffold, as was done here, is therefore an important step for reducing transmission line loss.

\section{Qubit-transmission line coupling}

Each qubit $Q_i$ is coupled to the transmission line via a tunable coupler $G_i$, based on a design in Ref.~\onlinecite{Chen2014}. This configuration is accurately modeled \cite{Geller2015} as a tunable inductance, with fixed inductances $L_g$ on each side of this tunable inductance (see Fig.~1b in the main text).

The effective mutual inductance between each qubit and the transmission line through the coupler is given by
\begin{equation}\label{M}
  M = \frac{L_g^2}{2L_g+L_w+L_T/\cos\delta}.
\end{equation}
Here $\delta$ is the phase across the Josephson junction that determines the equivalent inductance $L_T/\cos \delta$, and $L_w \approx 0.1$ nH represents the stray wiring inductance, which cannot be ignored when $L_T$ becomes very small (the stray wiring term does not appear in Ref. \onlinecite{Chen2014}).

\begin{figure}
  \begin{center}
  \includegraphics[width=3.0in]{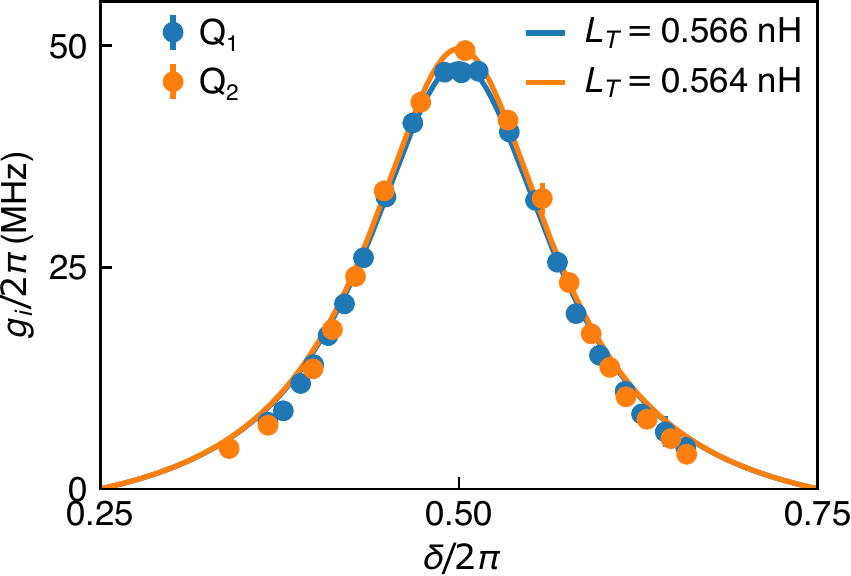}
  \caption{\label{figs4} Coupling strength versus coupler junction phase $\delta$. We measure the qubit spectrum at different coupler bias values, similar to Fig.~3a, and fit the spectrum to obtain the coupling strength $g_i$. The maximum coupling is about 47~MHz for $Q_1$ and 49~MHz for $Q_2$. Error bars are one standard deviation.}
  \end{center}
\end{figure}

In the harmonic limit and assuming weak coupling, the coupling between qubit $Q_i$ and the $n$th mode is
\begin{equation}\label{coupling_n}
  g_{i,n} = -\frac{M}{2} \, \sqrt{\frac{\omega_i \omega_n}{(L_g+L_J)(L_g+L_n)}}.
\end{equation}
We see that $g_{i,n} \propto \sqrt{\omega_n} \propto \sqrt{n}$, a well-known result for multimode coupling. The coupling depends on the control signals sent to the coupler $G_i$, and must be calibrated by fitting to measurements similar to those shown in Figs. 1c and 3a in the main text, involving typically 4 to 6 adjacent modes. It is experimentally more practical to approximate the coupling in these calibrations by a single value $g_i$, where as the mode numbers $n \sim 70$, the variation in $g_{i,n}$ with $n$ in the calibration is only about 2\%. Experiments reported here using these calibrations involve up to roughly 10 modes, for which the total variation in coupling is less than 5\%. These variations are small enough that this approximation is justified.  The calibration of the coupling $g_i$ as a function of the coupler phase is shown in Fig.~\ref{figs4} for each qubit $Q_i$.

The analytical result Eq.~(\ref{coupling_n}) agrees well with the experimental data, using $L_T = 0.566$~nH for $G_1$ and $L_T = 0.564$~nH for $G_2$. The comparison between this calculation and the measured coupling for both qubits is shown in Fig.~\ref{figs4}. Maximum coupling occurs at junction phase $\delta = \pi$, where we find $g_{i,{\rm max}}/2\pi \approx 47$~MHz for qubit frequencies near $5.8$~GHz. The coupling can be turned off by setting $\delta=\pi/2$, making $L_T/\cos \delta$ very large. We turn the couplers off when characterizing the qubits.

\subsection{Coupler-generated qubit frequency shift}
The tunable couplers used here ideally only change the qubit-transmission line coupling strength. However, changes in the coupler junction inductance $L_T$ affect the qubit resonance frequency, as can be seen from the circuit diagram in Fig.~1b in the main text. This is accounted for by including the coupler mutual inductance $M$, Eq.~(\ref{M}), in the calculation of the qubit frequency, through its effect on the qubit inductance $L_q$, which is given by
\begin{equation}\label{Lq}
  L_q = L_J +L_g - M.
\end{equation}
In the experiment, $\omega_n \sim \omega_i \approx \displaystyle \frac{1}{\sqrt{(L_g+L_J)C_q}}$, so we can use Eq.~(\ref{coupling_n}) to relate the mutual inductance to the coupling,
\begin{equation}\label{eq}
  M = -2g_i \sqrt{C_q(L_g + L_n)} (L_g+L_J).
\end{equation}
The qubit inductance is then given by
\begin{equation}\label{Lq_approx}
  L_q = (L_g +L_J)\left(1+2g_i \sqrt{C_q(L_g + L_n)}\right),
\end{equation}
so that the qubit frequency including the coupler is given by
\begin{eqnarray}\label{omega_q}
  \omega_i^\prime &=& \frac{1}{\sqrt{L_q C_q}} \\
  &=& \frac{1}{\sqrt{(L_g +L_J) C_q}}\frac{1}{\sqrt{1+2g_i \sqrt{C_q(L_g + L_n)}}} \\
  &\approx& \omega_{i} \left (1-g_i \sqrt{C_q(L_g + L_n)} \right ).
\end{eqnarray}
We therefore find that the qubit frequency is shifted by the coupler by an amount
\begin{equation}\label{shift}
  \Delta \omega_i = -g_i \omega_{i}  \sqrt{C_q(L_g + L_n)} = -g_i \sqrt{\frac{L_g+L_n}{L_g + L_J}}.
\end{equation}
Similarly, we can show that the transmission line's $n$th mode resonant frequency is shifted by
\begin{equation}\label{shiftn}
  \Delta \omega_n = -g_i \omega_{n}  \sqrt{C_n(L_g + L_J)} = -g_i \sqrt{\frac{L_g+L_J}{L_g + L_n}}.
\end{equation}
Because $L_n \gg L_J$, $\Delta \omega_i$ is much larger than $\Delta \omega_n$. According to Fig.~\ref{figs4}, with maximum coupling $g_{i,\rm{max}}/2\pi \approx 47$~MHz, the qubit frequency can be shifted by as much as $-200$ MHz by tuning the coupling from off to its maximum value. This frequency shift can be compensated by adjusting the qubit junction inductance $L_J$ accordingly, as was done in the measurements.

In the experiments shown in Fig.~3a and b in the main text, we bias $G_1$ to set its coupling to its maximum value, which changes the qubit frequency through Eq.~(\ref{shift}). At the same time, we adjust $Q_1$'s $Z$ bias, which changes the qubit junction inductance $L_J$. The net qubit frequency is determined by the combination of these two effects, and is calibrated by fitting the response in Fig.~3a. The data in Fig.~3c represent a special case of the data in Fig.~3b, where the qubit $Z$ bias is set to zero, keeping $L_J$ fixed, as represented by the horizontal line after the $\pi$ pulse in $Q_1$'s control sequence. However, the qubit frequency is still affected by the coupler. This impacts the itinerant photon capture efficiency, and must be accounted for in the simulations (see the Numerical Simulations section below).

In Fig.~3d, to optimize the itinerant photon capture, we adjust the qubit's $Z$ bias to change $L_J$ while tuning the coupling, such that the change of $L_J$ and $M$ cancel each other out, and the qubit frequency is fixed (ideally) during the photon emission and capture process. The two convolution pulses after the $\pi$ pulse in $Q_1$'s control sequence represent this counteracting $Z$ bias. In Fig.~4a in the main text, we similarly apply $Z$ bias pulses to both qubits while tuning the couplers, as shown by the control pulse sequences in the inset. These bias pulses not only counteract the frequency shift from the coupler, but also adjust the qubit frequencies to match each other, as the operating frequencies are not the same for the two qubits.

\section{CHSH Bell inequality}
After generating a Bell state using either the relay mode method or the shaped itinerant photon method, we perform the CHSH form of Bell inequality test. This is done by measuring $Q_1$ along either direction $a$, which is chosen to be the $x$ axis on the Bloch sphere (see inset to Fig.~3d), or along direction $a^\prime$, which is chosen to be the Bloch sphere $y$ axis. At the same time, we measure $Q_2$ along direction $b$ or $b^\prime$, where $b$ is on the Bloch sphere equator, rotated by an angle $\theta$ about the $z$ axis with respect to $a$, and $b^\prime$ is perpendicular to $b$. For each choice of axes $(q_1, q_2)$ (where $q_1$ can be $a$ or $a^\prime$, $q_2$ can be $b$ or $b^\prime$), we accumulate many measurements of the two qubits, and calculate the quantum correlation $E(q_1,q_2) = P_{gg}+P_{ee}-P_{ge}-P_{eg}$, where the subscript $ge$ for example means those measurement outcomes where $Q_1$ was measured to be in $|g\rangle$ along $q_1$ and $Q_2$ was measured to be in $|e\rangle$ along $q_2$. Given the set of four quantum correlators for a given angle $\theta$, we then define the CHSH correlation $S(\theta) = E(a,b) - E(a,b^\prime) + E(a^\prime,b) + E(a^\prime,b^\prime)$. The CHSH inequality states that $|S| \le 2$ for a classical system, while quantum physics predicts $|S| \le 2\sqrt{2}$. For an ideal Bell state, $S$ is maximized when $a \bot a^\prime$, $b \bot b^\prime$, and $a$ is at $\pi\pm 3\pi/4$ rad with respect to $b$, for the singlet and triplet Bell states respectively.

\section{Numerical Simulations}
\subsection{Multimode model simulation}
We performed extensive numerical simulations to better understand and calibrate the experiment. These simulations used the following rotating-frame qubit-multimode Hamiltonian:
\begin{equation}\label{H}
  H/\hbar = \sum_{i=1,2} \Delta\omega_{i} \sigma_{i}^\dag \sigma_{i} + \sum_{n=1}^{N} \left (n-\frac{N+1}{2} \right ) \omega_{\rm{FSR}} a_n^\dag a_n + \sum_{i=1,2} \sum_{n=1}^{N} g_{i,n} \left (\sigma_{i} a_n^\dag + \sigma_{i}^\dag a_n \right ),
\end{equation}
where $\sigma_i$ and $a_n$ are the annihilation operators for qubit $Q_i$ and photons in the $n$th standing wave mode, respectively, $\Delta\omega_{i}$ is the qubit frequency detuning in the rotating frame, and $N$ is the number of standing modes included in the simulation.

In Fig.~1c, we fit the qubit spectrum by solving for the eigenenergies of the qubit-multimode Hamiltonian, Eq.~(\ref{H}), including six transmission line standing modes and setting $g_{2,n}$ to zero.

In Fig.~2 in the main text, where the coupling is weak, we include five standing modes in the simulations, where the third mode relays the quantum state. The rotating frame frequency is chosen to be the third mode resonant frequency so that the modes are symmetrically distributed. The coupling is assumed to be turned on and off abruptly, as the coupler rise and fall times are significantly shorter than the swap time. Decoherence is taken into account using the Lindblad master equation. According to Ref. \onlinecite{Averin2016}, the effective dephasing time is enhanced by $\sqrt{2}$ when transferring the quantum state from one qubit to the other, because the dephasing noise at each qubit is uncorrelated. Taking this into account, we find that the simulation agrees well with the experiment. According to the simulations, more than half of the infidelity is attributed to dephasing noise. Simulations that take $T_2 = 10~\mu$s for both qubits give a state transfer process fidelity of $\mathcal{F}_1^p = 0.977$, a Bell state fidelity $\mathcal{F}_1^s = 0.983$ and a concurrence $\mathcal{C}_1 = 0.980$. The remaining 2 percent infidelity is attributed to energy dissipation and interference from adjacent modes.

For Fig.~3a, we fit the qubit spectrum by solving for the eigenenergies of the qubit-multimode Hamiltonian including now ten standing modes and setting $g_{2,n}$ to zero.

\subsection{Input-output theory simulation}
For the time domain experiments in Fig.~3 and Fig.~4 in the main text, the maximum coupling $g_{i,{\rm max}}$ becomes comparable to the free spectral range $\omega_{\rm FSR}$. To maintain the multimode model accuracy, the number of modes $N$ needed for the simulation (and thus the Hilbert space dimension) is so large that it becomes overly computer-intensive to complete the simulations. An alternative is to use input-output theory \cite{Gardiner1985,Cirac1997,Korotkov2011}, which treats the mode spectrum in the transmission line as continuous, and thus is well-suited for simulating the dynamics with large $g_{i,{\rm max}}/\omega_{\rm FSR}$.

First we consider the quantum ``ping-pong'' dynamics in Fig.~3c. According to the input-output theory~\cite{Gardiner1985}, the evolution of the qubit operator $\sigma_1$ follows
\begin{eqnarray}
 \frac{\rm{d} \sigma_1 (t)}{\rm{d}t} &=& -i\Delta\omega_1 (t) \sigma_1 (t) -\frac{\kappa_1(t)}{2} \sigma_1(t) + \sqrt{\kappa_1(t)} a_{\rm{in},1}(t), \label{io1}\\
 \sqrt{\kappa_1(t)} \sigma_{1}(t) &=& a_{\rm{in},1}(t) + a_{\rm{out},1}(t),\label{io2} \\
 a_{\rm{in},1}(t) &=& a_{\rm{out},1}(t-2T_\ell),\label{io3}
 \end{eqnarray}
where $\kappa_1$ is the qubit $Q_1$ energy decay rate to the transmission line, which can be calculated according to Fermi's golden rule:
 \begin{equation}\label{Fermi}
   \kappa_1 = \frac{2\pi}{\hbar} (\hbar g_1)^2\frac{1}{\hbar \omega_{\rm{FSR}}}.
 \end{equation}
The input and output field operators are $a_{\rm{in},1}$ and $a_{\rm{out},1}$, respectively. Note we have replaced the resonator annihilation operator by the qubit annihilation operator; this replacement is valid because we only consider situations with at most one excitation in the system.


We observe that as the coupling becomes strong, the finite rise and fall time of the control signal has to be taken into account. In the simulations, we assume the phase due to the external flux threaded through the coupler loop $\delta_{\rm ext}$ is proportional to the control pulse amplitude. The coupler junction phase $\delta$ is related to $\delta_{\rm ext}$ by \cite{Geller2015}
\begin{equation}\label{delta2ext}
  \delta_{\rm ext} = \delta + \frac{2L_g+L_w}{L_T}\sin\delta.
\end{equation}
The coupler is first biased with a DC current to give
\begin{equation}\label{deltaext0}
  \delta_{\rm ext} = \delta_{\rm off} = \pi/2 + \frac{2L_g+L_w}{L_T},
\end{equation}
where $\delta = \pi/2$ and $g_1 = 0$. We then use the high-speed control signal output of the DAC to rapidly tune the coupling $g_1$, combined with a separate DC current source via a bias tee mounted on the mixing chamber stage. The filter in the DAC output has a Gaussian spectrum, so that when we generate a rectangular output signal to set the coupling to its maximum value (where $\delta_{\rm ext}=\delta=\pi$), the actual output is a convolution of the filter Gaussian and the rectangular control signal. The external flux then changes as
\begin{equation}\label{deltaext}
  \delta_{\rm ext}(t) = (\pi-\delta_{\rm off}) \left (G(w_{\rm{FWHM}}, t) \circledast {\rm Rect}(\tau_g, t) \right )(t) + \delta_{\rm off},
\end{equation}
where $G(w_{\rm{FWHM}},t)$ is a Gaussian function with a full-width at half-maximum (FWHM) of $w_{\rm{FWHM}}$, and ${\rm Rect}(\tau,t)$ is a rectangle function with unit amplitude from 0 to $\tau_g$. We then solve Eq.~(\ref{delta2ext}) to obtain $\delta(t)$, and use this result in Eq.~(\ref{coupling_n}) to obtain $g_1(t)$. The energy decay rate $\kappa_1$ can then be calculated with Eq.~(\ref{Fermi}).

\begin{figure}
  \begin{center}
  \includegraphics[width=4.5in]{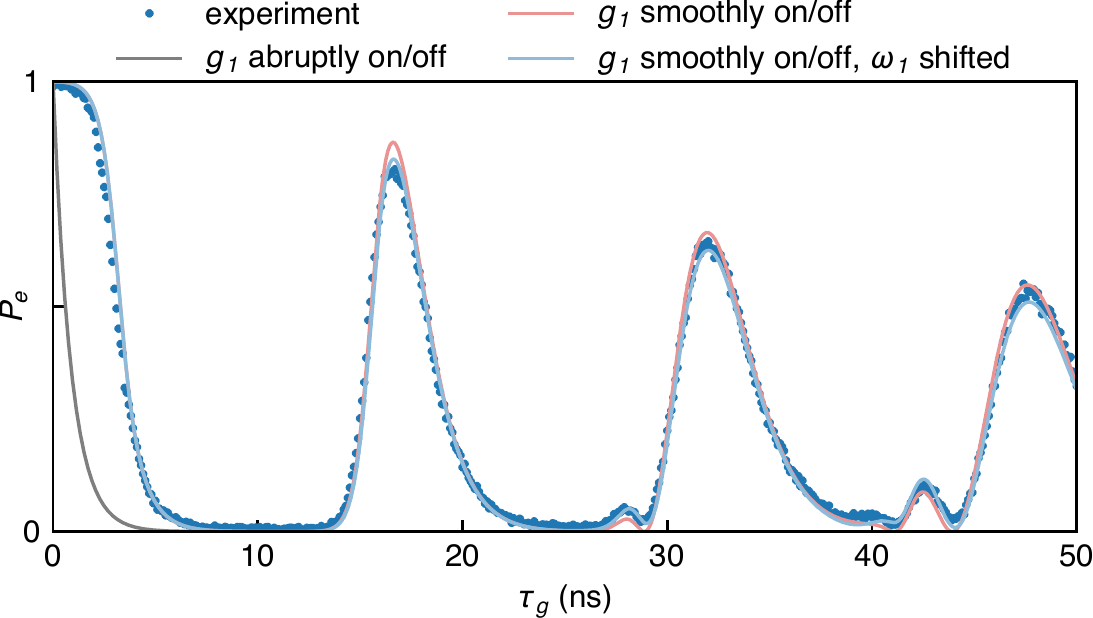}\\
  \caption{\label{figs5} Different simulations for the experiment shown in Fig.~3c. Simulation details are given in the text.}
  \end{center}
\end{figure}

In Fig.~\ref{figs5}, we compare the experimental data with different assumptions for the simulations. The light black line treats the coupling as switched abruptly between its on and off values, i.e. we assume $w_{\rm{FWHM}}=0$. We see that the qubit occupation decays exponentially in the simulation, and the recapture probability is limited to $\sim 54\%$, consistent with the calculations in Refs.~\onlinecite{Stobinska2009,Wang2011}. The light red line corresponds to setting $w_{\rm{FWHM}} = 2$ ns, which is determined by the bandwidth of the control signal output filter, and agrees well with the experimental data, except the photon recapture probability is higher. This is because the qubit frequency is shifted when tuning the coupling to the maximum, see Eq.~(\ref{shift}), not accounted for in this simulation. The light blue line takes the frequency shift into account and is in good agreement with the experiment.

In Fig.~3d, the coupler control signal changes as:
\begin{equation}\label{deltaext2}
  \delta_{\rm ext}(t) = (\pi-\delta_{\rm off}) \left \lbrace G(w_{\rm{FWHM}}, t) \circledast \left [ {\rm Rect}(\tau_g, t) + {\rm Rect}(\tau_g, t-\tau_w - \tau_g) \right ] \right \rbrace (t) + \delta_{\rm off}.
\end{equation}
In addition to the Gaussian filter, we program the control signal output to adjust $w_{\rm{FWHM}}$ to 3 ns to optimize the photon catch probability, and compensate the qubit frequency shift with the qubit $Z$ bias pulse. This frequency compensation is assumed to be perfect in the simulation, i.e., $\Delta\omega_1=0$.

Here we model the state transfer process in Fig.~4a using input-output theory~\cite{Gardiner1985}:
\begin{eqnarray}
  \frac{\rm{d} \sigma_1}{\rm{d}t} &=& -i\Delta\omega_1 \sigma_1 -\frac{\kappa_1(t)}{2} \sigma_1 + \sqrt{\kappa_1(t)} a_{\rm{in},1}(t), \label{2io1}\\
  \frac{\rm{d} \sigma_2}{\rm{d}t} &=& -i\Delta\omega_2 \sigma_2 -\frac{\kappa_2(t)}{2} \sigma_2 + \sqrt{\kappa_2(t)} a_{\rm{in},2}(t), \label{2io2}\\
  \sqrt{\kappa_1(t)} \sigma_{1}(t) &=& a_{\rm{in},1}(t) + a_{\rm{out},1}(t),\label{2io3} \\
  \sqrt{\kappa_2(t)} \sigma_{2}(t) &=& a_{\rm{in},2}(t) + a_{\rm{out},2}(t),\label{2io4} \\
  a_{\rm{in},1}(t) &=& a_{\rm{out},2}(t-T_\ell),\label{2io5} \\
  a_{\rm{in},2}(t) &=& a_{\rm{out},1}(t-T_\ell).\label{2io6}
\end{eqnarray}

The time evolution of the decay rates $\kappa_i (t)$ are calculated as mentioned above for single qubit ``ping-pong'' with itinerant photons. The qubit frequency shifts are assumed to be perfectly compensated in the simulation, so we take $\Delta \omega_i = 0$. The simulated emission agrees very well with the $Q_1$ data, and the simulated capture agrees with the $Q_2$ data at the beginning, but reaches a higher maximum capture probability than the experiment. According to ref.~\onlinecite{Sete2015}, the state transfer protocol is robust against control pulse imperfections, but is sensitive to qubit frequency mismatch. The discrepancy between the simulation and the experiment is likely due to the frequency mismatch between the two qubits. Note the state transfer process fidelity is not affected by changes in the transmission line length $\ell$ in this simulation, unless the channel decoherence is taken into account.

\begin{figure}
  \begin{center}
  \includegraphics[width=4.5in]{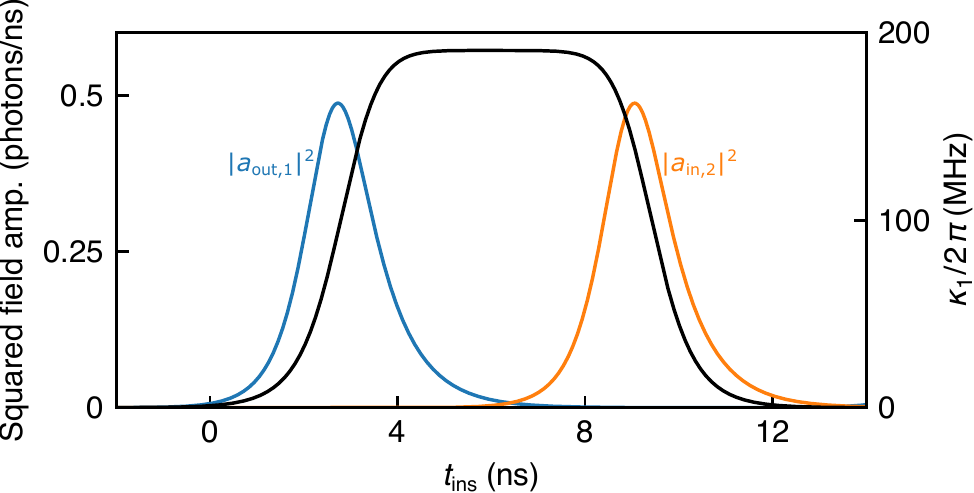}\\
  \caption{\label{envelope}Traveling photon envelope estimated from simulations. The horizontal axis is the instantaneous time $t_{\rm ins}$ of the dynamic evolution, calculated for a control pulse width set to $t=12.2$~ns for optimized state transfer. The blue curve shows the emitted photon envelope $|a_{\rm{out},1}|^2$; the orange curve shows the captured photon envelope $|a_{\rm{in},2}|^2$. The black curve (right axis) shows the decay rate $\kappa_1/2\pi$ for qubit $Q_1$; qubit $Q_2$ is very similar.
  }
  \end{center}
\end{figure}

In the experiments in Refs. \onlinecite{Kurpiers2018, Axline2018, Campagne2018}, a circulator was interposed in the transmission line connecting the two communication nodes, eliminating reflections and at the same time providing a means to probe the emitted photon waveform, allowing tune-up of the emission profile to achieve the desired symmetric envelope. In our itinerant photon experiment, we have no direct means to probe the emitted photon envelope. However, the emitted and captured photon envelope can be estimated from input-output theory. In Fig.~\ref{envelope} we show $|a_{\rm{out},1}|^2$ and $|a_{\rm{in},2}|^2$ calculated from the simulation results shown in Fig.~4a in the main text, these results being close to the experimental data. We see that the emitted photon envelope is relatively symmetric, even with the simple coupler control pulse used in the experiment. This symmetry is the key reason that we are able to achieve such high-fidelity state transfers using the itinerant photon method.

\section{Optimized itinerant photon catch for $Q_2$}
\begin{figure}
  \begin{center}
  \includegraphics[width=4.5in]{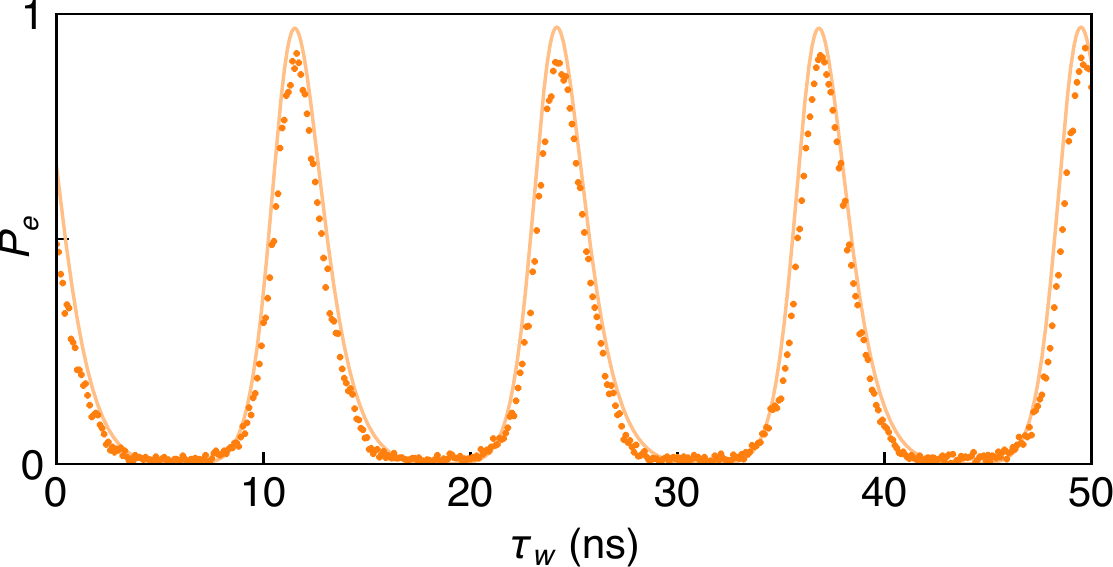}\\
  \caption{\label{Q2catch} Optimized itinerant photon catch process for qubit $Q_2$, analogous to Fig.~3d in the main text with a slight adjustment to the control pulse optimization. The capture probability is found to be $0.917 \pm 0.006$. Solid line is simulation.}
  \end{center}
\end{figure}

In Fig.~3d in the main text, we show the data for optimizing qubit $Q_1$'s itinerant photon ``catch'' process. Here we show the analogous data for qubit $Q_2$, see Fig.~\ref{Q2catch}. The maximum photon catch probability is found to be $0.917 \pm 0.006$.

\section{Quantum state and process tomography}
Quantum state tomography is performed after the Bell state preparation by applying the tomography gates $\lbrace I, ~R_x^{\pi/2}, ~R_y^{\pi/2}\rbrace$ to each qubit and reading out both qubits simultaneously. The density matrix is then reconstructed using linear inversion. The density matrix is validated to guarantee that it is Hermitian and positive with unit trace. In the experiment, the $R_x^{\pi/2}$ and $R_y^{\pi/2}$ tomography pulses are rotated with a calibrated angle about the Bloch sphere equator to cancel the qubit dynamical phase accumulated during state preparation.

The quantum process tomography for the state transfer is carried out by preparing $Q_1$ in the input states $\lbrace |g\rangle,~(|g\rangle-i|e\rangle)/\sqrt{2}, ~(|g\rangle+|e\rangle)/\sqrt{2},~|e\rangle \rbrace$, then performing the quantum state transfer process. The corresponding outcome density matrix in $Q_2$ is measured using quantum state tomography. The process matrix is obtained using the least squares approximation from these input and outcome states.  The process matrix is validated to guarantee it is Hermitian, positive and trace-preserving. We note that in quantum optics, a non-trace-preserving process matrix is typically used to characterize the quantum state transfer, accounting for loss in the transmission channel. Here energy dissipation in the channel is indistinguishable from the ground state transfer on the receiver end; it is therefore natural to use a trace-preserving process matrix to characterize the state transfer, although the dissipation in the transmission line is negligible.

\clearpage

\bibliography{bibliography}
\bibliographystyle{naturemag}